\documentclass[a4paper,11pt]{article}
\pdfoutput=1 
\usepackage{jinstpub} 
\usepackage{url}
\usepackage{subcaption}

\title{\boldmath Geant4 Silver Anniversary: 25 years enabling scientific production}

\author[a,1]{T. Basaglia\note{Retired},}
\author[b]{Z. W. Bell,}
\author[c,d]{D. D'Agostino,}
\author[b]{P. V. Dressendorfer,}
\author[d,2]{M. G.  Pia,\note{Corresponding author.}}
\author[e]{E. Ronchieri}

\affiliation[a]{CERN, \\
CH-1211 Geneva 23, Switzerland}
\affiliation[b]{ IEEE, \\
445 Hoes Lane, Piscataway, NJ 08854, USA}
\affiliation[c]{DIBRIS, Univ. of Genova, \\Via Dodecaneso 35, 16146 Genova, Italy}
\affiliation[d]{INFN Sezione di Genova, \\Via Dodecaneso 33, 16146 Genova, Italy}
\affiliation[e]{INFN CNAF,\\Viale Carlo Berti Pichat 6, 40127 Bologna, Italy}

\emailAdd{mariagrazia.pia@ge.infn.it}

\abstract{ This paper summarizes Geant4 contribution to scientific research over
the past 25 years through a scientometric analysis of the results with which it has been
associated. The scientometric data collected from scholarly literature and
databases are evaluated with methods pertaining to econometrics and ecology to
quantify relevant traits, diversity and disparity in their scientific and
geographic distributions, and to identify statistically significant trends. The
analysis reviews the contribution of Geant4 to the field - experimental particle
physics - that originally motivated its development and highlights its role in
other research domains including nuclear physics and engineering, astrophysics
and space science, biomedical physics, archaeology and the cultural heritage.}


\proceeding{16$^{\text{th}}$ Topical Seminar on Innovative Particle and Radiation Detectors \\
  25-29 September 2023\\
  Siena, Italy}

\begin{document}
\maketitle
\flushbottom

\section{Introduction}
\label{sec:intro}

Geant4 \cite{g4nim, g4tns, g4nim2} turned 25 years old in 2023.
Its main reference \cite{g4nim}, published in 2003, has achieved more than
16000 citations in the Web of Science\textsuperscript{\texttrademark}
\cite{wos} by September 2023, making it the most cited publication in Particle and
Fields Physics, Nuclear Physics, Nuclear Science and Technology, and Instruments
and Instrumentation journals at the time of writing this paper. 
The citations come from a wide variety of research areas, which demonstrate the
broad scope of scientific results that Geant4 has enabled.

This paper briefly summarizes some relevant characteristics of the research
production associated with Geant4 through the analysis of the citations of its
main reference \cite{g4nim} in the scholarly literature.
Along with some basic traits of these publications pertaining to descriptive
statistics, it reports a scientometric investigation of Geant4 citations by
means of statistical methods derived from econometrics and quantitative ecology,
which highlight characteristics such as fairness and diversity, and their
trends, in the scientific production enabled by Geant4.

\section{Brief chronicle of Geant4 birth}

Geant4 is an object oriented toolkit for the simulation of the passage of 
particles through matter.
Its development was initially motivated by the requirements of physics
experiments at high energy hadron colliders under construction in
the last decade of the 20$^{th}$ century.
Due to intrinsic limitations \cite{P58}, GEANT 3 could not address their needs
of functionality, extensibility, flexibility and long term maintainability.

Early considerations about developing an object oriented Monte Carlo system for
particle transport targeted to this experimental domain date back to 1993
\cite{takaiwa_1993, giani_1993}.
A letter of intent \cite{intent} and a following formal research proposal
\cite{P58} originating from these preliminary explorations were submitted to the
CERN Detector R\&D Committee, which led to the approval of the RD44 project 
in 1994 to develop Geant4.
RD44 had the mandate of creating a detector simulation toolkit for the next
generation of high energy physics experiments, namely the experiments at the LHC
(Large Hadron Collider), which were under construction at the time of its endorsement
in the CERN research program.

RD44 produced the first $\alpha$-version of Geant4  \cite{rd44-97} in April 1997, whose
functionality was comparable to that of GEANT 3, and the 
first $\beta$-version in July 1998. 
Geant4 was first released on 15 December 1998 \cite{rd44-98}; since then, new
versions of the code have been issued once or twice a year.

Although originally addressed to high energy physics experiments, Geant4
encompasses functionality relevant to other physics domains.
The adoption of the object oriented technology and the sound software design
conceived in the RD44 project are the key factors of Geant4 multidisciplinary
capabilities.
 
\section{Scientometric analysis}

The scientometric investigation concerns the publications citing Geant4 main
reference \cite{g4nim}; these papers are representative of the scientific
production that Geant4 has enabled.
For comparison with Geant4, similar assessments are performed for the main
references of other highly cited physics software systems
and for the publication of the discovery of the Higgs boson at the
LHC \cite{higgs_atlas, higgs_cms}, which constitutes a landmark in the physics
context that motivated the development of Geant4.

The scientometric research is based on data collected from the Web of
Science\textsuperscript{\texttrademark} (WoS) \cite{wos} and has access
to the WoS publication records since 1990.
The survey encompasses all main types of scholarly literature indexed in the
WoS: regular articles, reviews, letters etc. 
Conference papers are included in the analysis only if tagged as articles in the
WoS, i.e. papers generally published in scholarly journals. This constraint
ensures the access to all their associated scientometric data required in the
course of the analysis.


The first stage of the analysis highlights the main characteristics of the
literature citing Geant4 through a set of distributions pertaining to
descriptive statistics.
Further elaboration of the scientometric data exploit statistical methods
pertaining to econometrics and ecology to investigate the fairness of the
geographical apportionment and the diversity of the scientific production
associated with Geant4 and the other target references.
These assessments are complemented by statistical inference methods that
identify trends in the evolution of the observables; the Mann-Kendall test
\cite{mann_1945, kendall_1948} and the Cox-Stuart \cite{cox_1955} test are
applied for this purpose.
The significance level for the rejection of the null hypothesis is set at 0.01
for both tests, unless otherwise specified.

Additionally, the scientometric analysis examines the role of Geant4
in developments of industrial interest through an assessment of patents
associated with it.

The analysis uses the R software system \cite{R}. 




\section{General features of Geant4 citations}

The scientometric research reported in this paper is based on the analysis of
the citations of reference \cite{g4nim} registered in the Web of Science.
This paper is the first publication about Geant4 in a refereed
journal; previous information about it only appeared in institutional
documents and brief conference communications, which lack 
coverage in the WoS for extensive scientometric analysis.

The citations of Geant4 reference \cite{g4nim} amounted to 16023 on 14 September
2023.
The WoS identifies it as the most cited paper among more than 1.4 million
publications classified in the categories the hosting journal belongs to: Particle and
Fields Physics (encompassing 436176 publications in total),
Nuclear Physics (335328 publications), Nuclear Science and Technology (380632
publications), and Instruments and Instrumentation (696376 publications).

\begin{figure}[htbp] 
\centering 
\includegraphics[width=.7\textwidth,trim=0 0 8 20,clip]{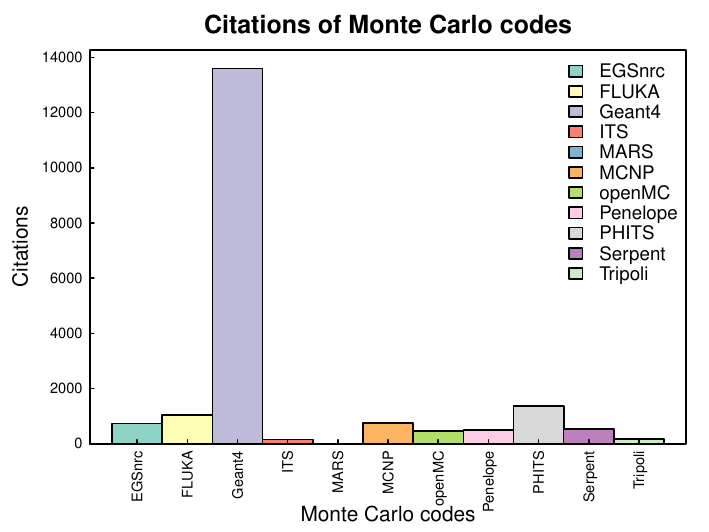} %
\caption{Citations of the reference papers associated with Monte Carlo particle
transport codes, as reported in the Web of Science.}
\label{fig:mcCitations} 
\end{figure}

Geant4 stands out among Monte Carlo particle transport codes commonly used in
experimental particle and nuclear physics with respect to the number of citations
received by the associated reference papers.
The distribution of the citations related to these codes 
(EGSnrc\cite{egsnrcJournal}, 
FLUKA \cite{fluka11}, 
ITS \cite{its3},
MARS \cite{mars},
MCNP 6 \cite{mcnp6},
OpenMC \cite{openmc1},
PENELOPE \cite{penelope1995},
PHITS \cite{phits2002},
Serpent \cite{serpent2013},
TRIPOLI-4\textsuperscript{\textregistered} \cite{tripoli4})
is illustrated in figure \ref{fig:mcCitations}.
Caution should be exercised at drawing quantitative conclusions from this plot
for various reasons: the citation of the references of Geant4 and of other Monte
Carlo codes is omitted in a large number of papers that mention them and use
them to produce the scientific results they document \cite{basaglia_2017};
additionally, those Monte Carlo transport codes 
not associated with a
journal publication,  consequently not indexed in the WoS,
are not represented in figure \ref{fig:mcCitations}.

\begin{figure}[htbp]
\centering 
\begin{subfigure}{0.33\textwidth}
    \centering
\includegraphics[width=\textwidth,trim=2 0 7 20,clip]{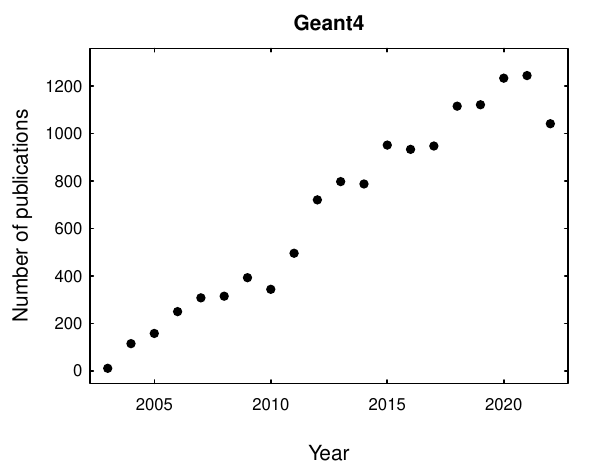}
   \caption{\label{fig:basic_ci} Citations}
    \end{subfigure}
\centering 
\begin{subfigure}{0.33\textwidth}
    \centering
\includegraphics[width=\textwidth,trim=0 0 7 20,clip]{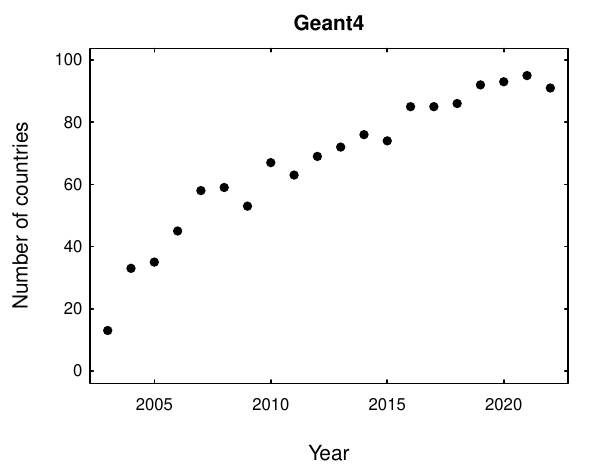}
   \caption{\label{fig:basic_co} Countries}
    \end{subfigure}
\centering 
\begin{subfigure}{0.329\textwidth}
    \centering
\includegraphics[width=\textwidth,trim=0 0 7 20,clip]{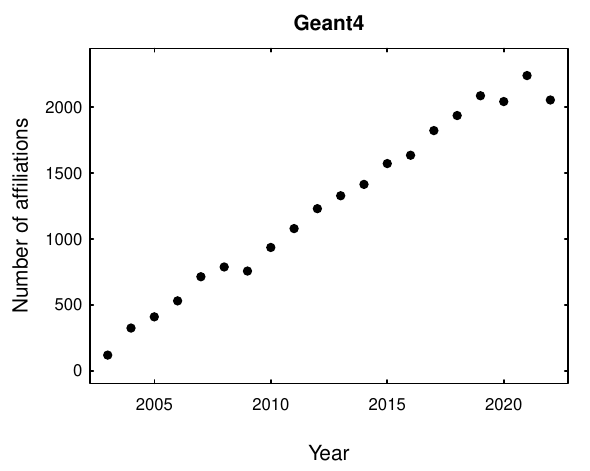}
   \caption{\label{fig:basic_af} Affiliations}
    \end{subfigure}
\caption{From left to right: number of publications citing Geant4 main reference paper, 
number of countries and number of affiliations appearing in the citing publications, as a function
of the publication year.}
\label{fig:basicG4}
\end{figure}

The number of citations of \cite{g4nim}, the number of countries and the number of
affiliations in the citing publications have been steadily growing, as is
illustrated in figure \ref{fig:basicG4} and objectively confirmed
by the Mann-Kendall and Cox-Stuart trend tests.
These tests reject the null hypothesis
of absence of any trend in the data in favour of the alternative hypothesis of a
growing trend.

\begin{figure}[htbp]
\centering 
\begin{subfigure}{0.33\textwidth}
    \centering
\includegraphics[width=\textwidth,trim=0 0 7 20,clip]{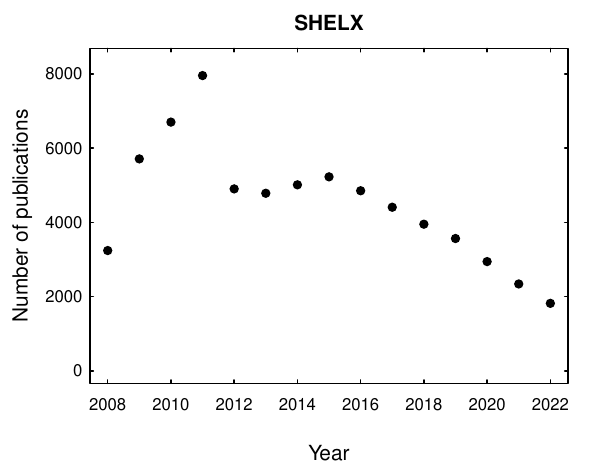}
    \caption{\label{fig:nPapers_shelx} SHELX}
    \end{subfigure}
\begin{subfigure}{0.33\textwidth}
    \centering
\includegraphics[width=\textwidth,trim=0 0 7 20,clip]{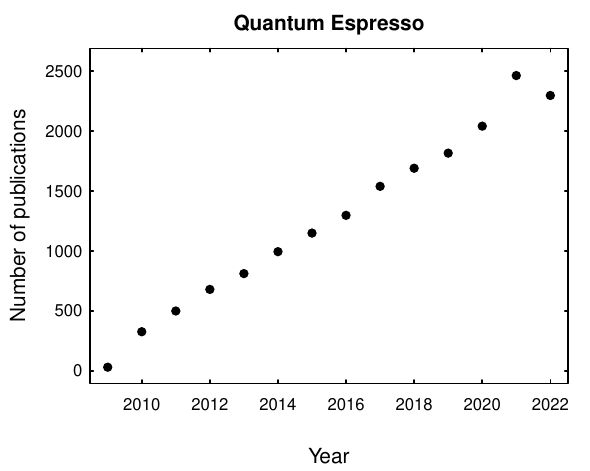}
    \caption{\label{fig:nPapers_qe} Quantum Espresso}
    \end{subfigure}
\begin{subfigure}{0.329\textwidth}
    \centering
\includegraphics[width=\textwidth,trim=0 0 7 20,clip]{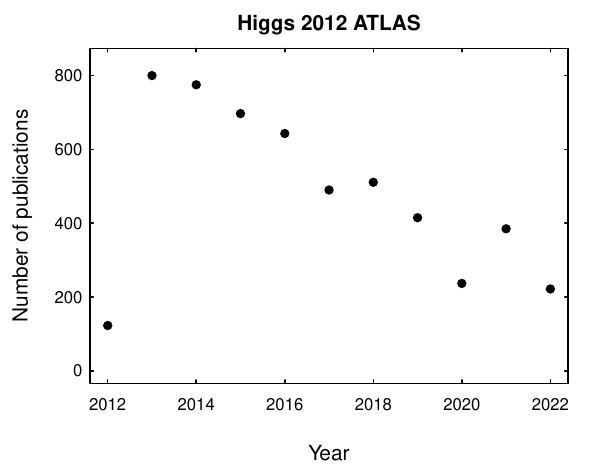}
    \caption{\label{fig:nPapers_higgs} Higgs boson observation}
    \end{subfigure}
\caption{From left to right: number of publications citing the reference papers
of the SHELX and Quantum Espresso codes, number of citations of the Higgs boson observation
by the ATLAS experiment, as a function of the publication year.}
\label{fig:nPapersOther}
\end{figure}

The longevity of Geant4 citations is remarkable in the context of
scientific software: for instance, one can observe in figure
\ref{fig:nPapersOther} a rapid decline of the citations of the SHELX
\cite{shelx} software system, whose reference paper published in 2008 has
collected more than 76000 citations.
The citations of other currently popular codes exhibiting a growing trend , e.g. 
Quantum Espresso \cite{qespresso}, span a shorter time range than Geant4
lifetime.
Note that the fast decrease of the citations of the observation of
the Higgs boson after its discovery is consistent with a previous scientometric 
assessment \cite{pia_2014} of the relatively short mean life of the citations of 
particle physics discoveries.
This drop could be related to the citation pattern concerning
foundational discoveries mentioned in \cite{vannoorden_2014},
which quickly become so familiar that they do not need a citation.

\begin{figure}[htbp]
\centering 
\begin{subfigure}{0.33\textwidth}
    \centering
\includegraphics[width=\textwidth,trim=0 0 8 20,clip]{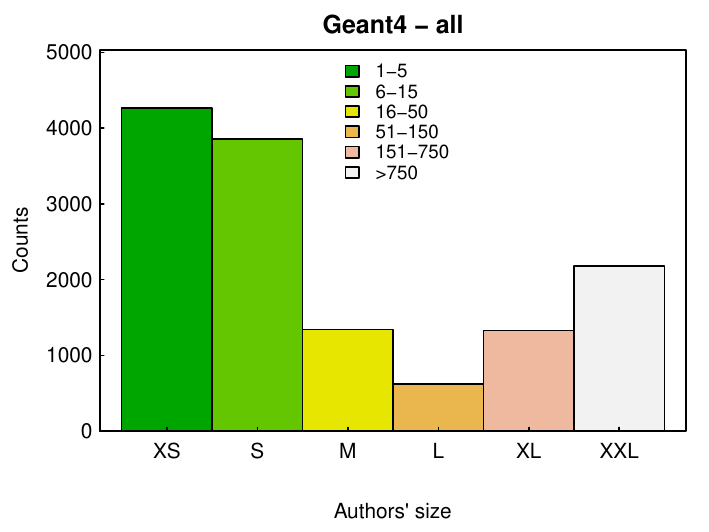}
  \caption{\label{fig:nAuthors_g4nim} Geant4}
    \end{subfigure}
\begin{subfigure}{0.33\textwidth}
    \centering
\includegraphics[width=\textwidth,trim=0 0 9 20,clip]{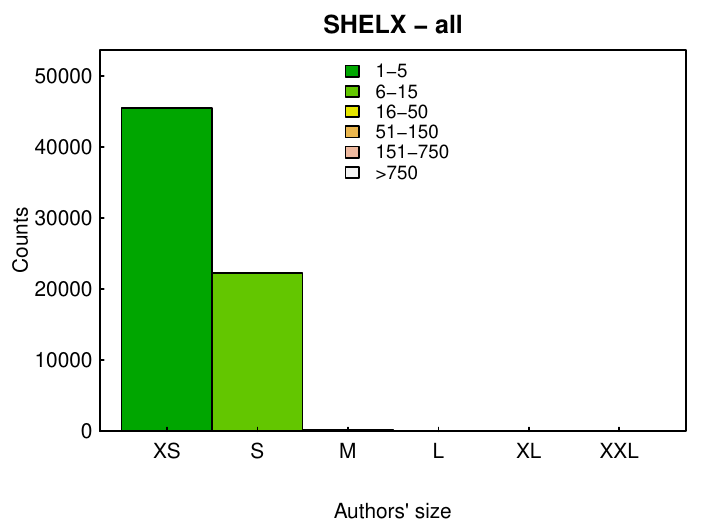}
  \caption{\label{fig:nAuthors_shelx} SHELX}
    \end{subfigure}
\begin{subfigure}{0.329\textwidth}
    \centering
\includegraphics[width=\textwidth,trim=0 0 0 20,clip]{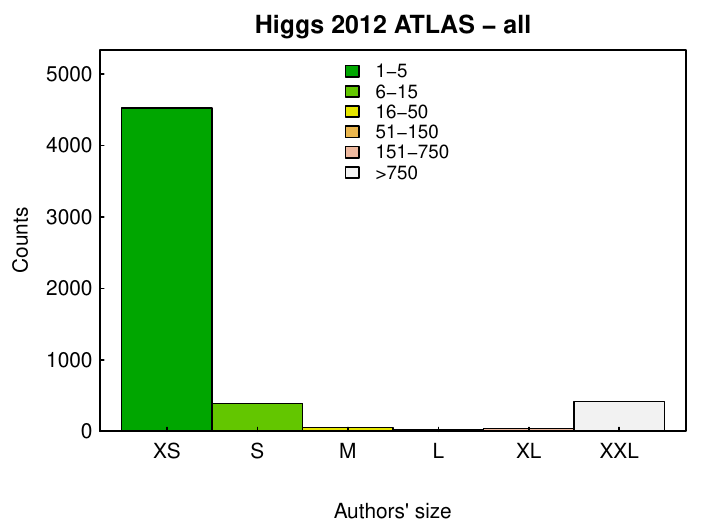}
  \caption{\label{fig:nAuthors_higgs} Higgs boson observation}
    \end{subfigure}
\caption{From left to right: size of the authors' list in the publications citing Geant4 and SHELX
reference papers, and the Higgs boson observation by the ATLAS experiment; the number of authors 
is binned according to the colour codes documented in the legend.}
\label{fig:nAuthors}
\end{figure}

A distinctive characteristic of Geant4 with respect to other highly cited
physics software systems is its ability to satisfy the requirements of different sizes
of scientific projects, from those pursued by small research teams to
large-scale collaborations enrolling thousands of members, as is 
illustrated in figure \ref{fig:nAuthors}.
It is interesting to note in figure \ref{fig:nAuthors_higgs} that the citations
of the observation of Higgs boson, which is the product of huge experimental
collaborations, mainly derive from small authors' teams (presumably theoretical
physicists) and only marginally from large collaborations.

\section{Geographical distribution}

The citations of Geant4 reference \cite{g4nim} derive from 114 countries.
Among them, there are large differences in population, wealth and scientific
research facilities.
The scientometric analysis investigated the degree of inequality in the
geographical origin of the authors of the citing papers by means of 
econometric methods.


Inequality measures \cite{ineq_handbook} quantify the degree of non-uniformity
of the distribution of a characteristic within a data set.
They are widely used  especially in evaluating the distribution of income.

\begin{figure}[htbp]
\centering 
\begin{subfigure}{0.45\textwidth}
    \centering
\includegraphics[width=\textwidth,trim=0 0 0 20,clip]{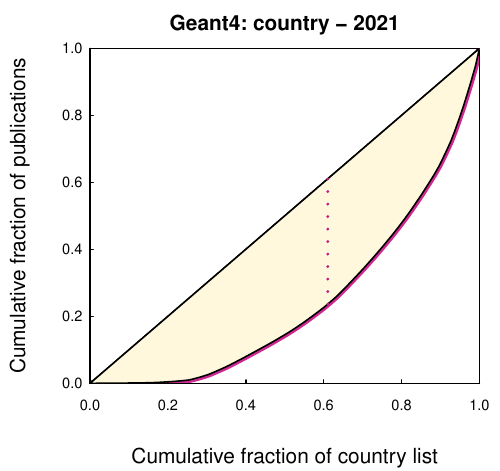}
  \caption{\label{fig:lorenz2_g4nim} Geant4}
    \end{subfigure}
\begin{subfigure}{0.45\textwidth}
    \centering
\includegraphics[width=\textwidth,trim=0 0 0 20,clip]{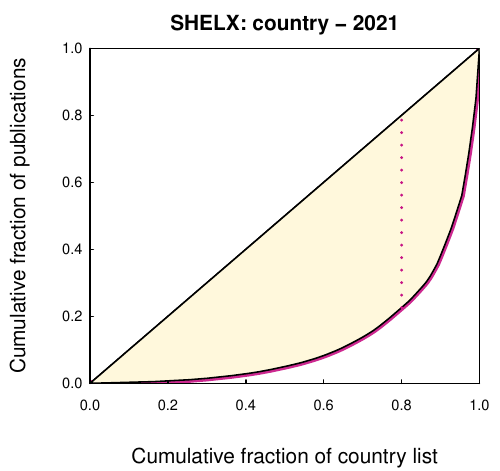}
 \caption{\label{fig:lorenz2_shelx} SHELX}
    \end{subfigure}
\caption{Graphical illustration of inequality in the geographical origin of the 
citations of Geant4 and SHELX in 2021: the coloured area between the diagonal
line and the Lorenz curve (solid red line) represents half the Gini index, while
the dotted red line represents the Pietra index. The larger the coloured area,
and the longer the dotted segment, the larger the inequality in the countries
represented in the authors' list of the citing papers. }
\label{fig:lorenz2}
\end{figure}

The Gini index \cite{gini_1912} is the most widely used measure of inequality in
economics; it indicates the extent to which a distribution deviates from a
situation of perfect equality.
It is calculated as 
\begin{equation}
G = 2 \, \int_{0}^{1}(x - L(x)) \,dx
\label{eq:gini}
\end{equation} 
where $L(x)$ is the Lorenz curve, which represents the cumulative share of total
values of the analyzed variable (e.g., income) against the cumulative proportion
of elements of the population being analyzed.
The diagonal line shown in figure \ref{fig:lorenz2} 
corresponds to perfect equality in graphical representations 
of the Lorenz curve.
A Gini index larger than 0.5 is generally perceived in economics as a measure of
unfair income distribution.

Other common measures of inequality are the Pietra index \cite{pietra_1915}, 
also related to the Lorenz curve, defined as 
\begin{equation}
P = max(x - L(x))
\label{eq:pietra}
\end{equation}
and the Atkinson index \cite{atkinson_1970}.

\begin{figure}[htbp]
\centering 
\begin{subfigure}{0.49\textwidth}
    \centering
\includegraphics[width=\textwidth,trim=0 0 0 20,clip]{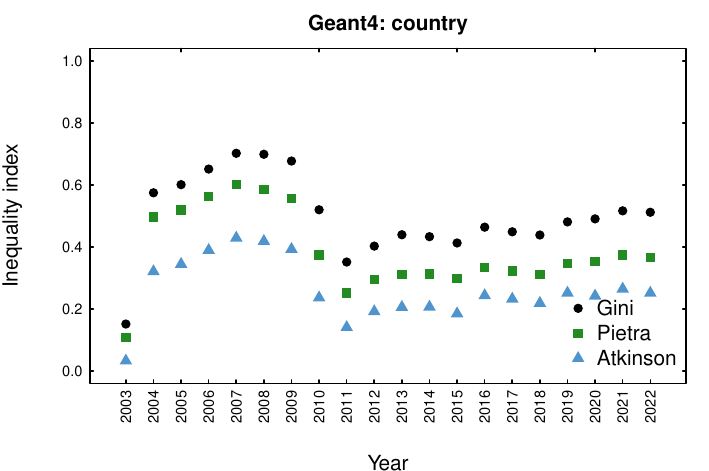}
  \caption{\label{fig:ineq3_g4nim} Geant4}
    \end{subfigure}
\begin{subfigure}{0.49\textwidth}
    \centering
\includegraphics[width=\textwidth,trim=0 0 0 20,clip]{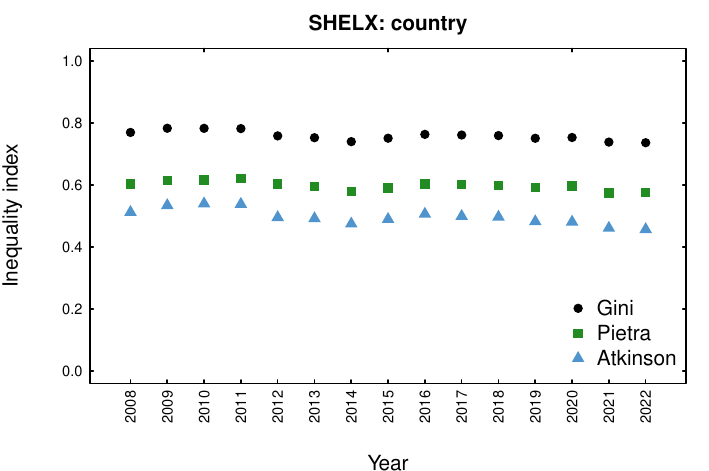}
 \caption{\label{fig:ineq3_shelx} SHELX}
    \end{subfigure}
\caption{Inequality in the distribution of the countries represented in the authors' list 
of the publications citing Geant4 and SHELX as a function of the publication year: Gini, Pietra and Atkinson indices.}
\label{fig:ineq3}
\end{figure}

Figure \ref{fig:lorenz2} shows a graphical representation of the Gini and Pietra
indices associated with the distribution of countries in the citations of Geant4
and SHELX in 2021.
In each panel, 
the  coloured area is the integral in equation \ref{eq:gini}
and the dotted line  represents the Pietra index.
Together they qualitatively indicate lower inequality in the geographical 
origin of the citations of Geant4 with respect to SHELX.
The measured geographical inequality is shown for the two codes in figure \ref{fig:ineq3}
as a function of time.
One can observe the similar time profiles characterizing the Gini, Pietra and Atkinson 
indices, despite the different values corresponding to their specific definitions of inequality,
and the lower geographical inequality exhibited by Geant4 citations with respect 
to those of SHELX over the whole period.
The drop of the three indices related to Geant4 in the years between 2007 and
2011 was influenced by the wider adoption of Geant4 for detector simulation by
international collaborations.

\section{Diversity of research areas}

The object oriented design of Geant4 and its rich functionality enable a wide
variety of applications in multidisciplinary domains.
The publications that cite Geant4 represent more than 60 research areas, defined
according to the criteria adopted in the WoS.
Among them, there are domains where Monte Carlo particle transport codes have been 
traditionally used for a long time, such as Physics,
Nuclear Science and Technology,
Astronomy and Astrophysics,
Instruments and Instrumentation,
Radiology, Nuclear Medicine and Medical Imaging; 
one can also identify less conventional application areas, such as Geology,
Archaeology, Food Science, Meteorology and many more.
We investigated the diversity in the research areas 
associated with Geant4 citations, drawing concepts and methods from
the domain of ecology.

Biodiversity measures the richness and the complexity of a community,
taking into account the number of species it hosts and their abundance;
it is related to the concept of entropy in information theory.

Hill indices, also known as Hill numbers \cite{hill_1973}, have recently
achieved a wide consensus as measures of diversity \cite{roswell_2021}: they
combine the concepts of richness, evenness and dominance, and their metrological
properties allow the comparisons of different systems.
They are defined as:
\begin{equation}
^qD = (\sum_{i=1}^S  p_i^q)^{\large\frac{1}{1-q}} 
\end{equation}
where $S$ is the number of species, $p_i$ is the proportional abundance of
species $i$ in the sample. 
The parameter $q$ is defined as the order of the index and is related to the
sensitivity to rare species in the corresponding measure of diversity.
Hill numbers can be interpreted as effective numbers of species, i.e. as the number of
equally abundant species necessary to produce the observed value of diversity.

Several traditional diversity indices \cite{magurran_2010}, such species richness, 
Shannon entropy \cite{shannon_1948} and Simpson index \cite{simpson_1949}, 
can be derived from Hill numbers:
the Hill index of order 0 corresponds to the number of species $S$;
the Hill index of order 1 is defined as the limit
\begin{equation}
^1D = \lim_{q \to 1} (^qD) = \exp(-\sum_{i=1}^S p_i \ln p_i)
\end{equation}
and corresponds to the exponential of the Shannon diversity index;
the Hill index of order 2 is the reciprocal of Simpson index.

\begin{figure}[htbp] 
\centering 
\includegraphics[width=.7\textwidth,trim=0 0 0 20,clip]{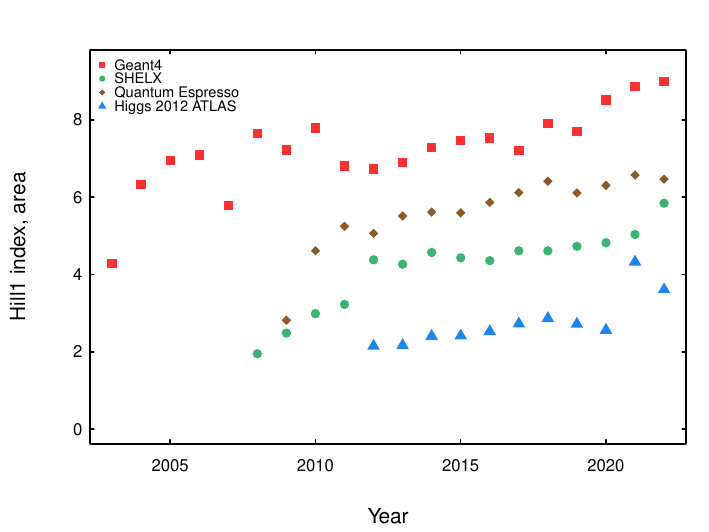} %
\caption{Hill index of order 1 as a function of time, 
representing the diversity of research areas  in the citations of
the reference papers of Geant4 and of other physics software system, and in
the citations of the observation of the Higgs boson at the LHC.}
\label{fig:hill1} 
\end{figure}

A sample of the analysis of diversity of research areas in the scientometric data 
is illustrated in figure \ref{fig:hill1}.
The plot shows the time profile of the Hill index of order 1 calculated over the
sets of research areas associated with the citations of Geant4, SHELX and
Quantum Espresso reference papers, and of the discovery of the Higgs boson.
Geant4 citations exhibit the largest diversity.
Both Mann-Kendall and Cox-Stuart
trend tests reject the null hypothesis in favour of the alternative hypothesis of a
growing trend in the data distribution related to Geant4; their response is not 
univocal about the data distributions related to SHELX and Quantum Espresso.

\section{Patents}

\begin{figure}[htbp] 
\centering 
 
\includegraphics[width=.7\textwidth,trim=0 0 0 20,clip]{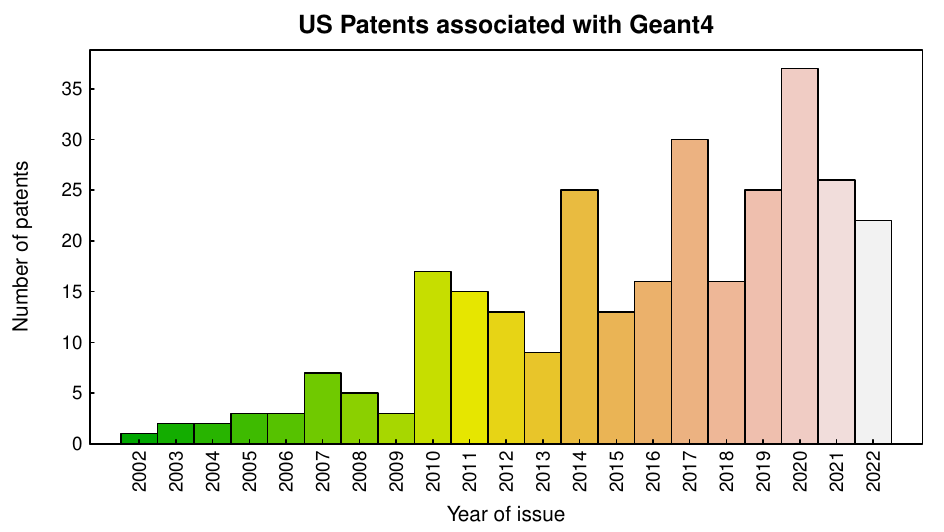} %
\caption{Patents associated with Geant4 issued by  the 
United States Patent and Trademark Office (USPTO).}
\label{fig:patents} 
\end{figure}

Geant4 has enabled many application developments of industrial and commercial
relevance. 
Several hundred patents are associated with it; the data are shown
in figure \ref{fig:patents}, which has the number of patents related to
Geant4 issued by the United States Patent and Trademark Office (USPTO) 
between 2002 and 2022.

The increasing relevance of Geant4 as a supporting document
for patents, which
one can qualitatively observe in figure \ref{fig:patents}, is confirmed by the outcome of the
Mann-Kendall and Cox-Stuart trend tests: both tests reject the null hypothesis
of absence of any trend in the period subject to investigation with 0.01
significance, in favour of the alternative hypothesis of a growing trend.

\section{Conclusion}

Our data show that Geant4 has a multidisciplinary nature, providing functionality for the simulation
of experimental scenarios in many different scientific fields. 
Its foundation lies in the object oriented design defined by RD44, which 
allows the user to understand, extend and customise the toolkit.

Geant4 use has been continuously growing for the past 25 years in an increasing
number of institutes and countries.
At the time of its silver anniversary, its main reference has established a
record of citations both in fundamental physics and in technological research
domains.
The scientometric analysis summarized in this paper has highlighted 
the diversity of research areas where Geant4 has enabled the production of scientific 
results and the fair geographical distribution of its use, as well as its contribution 
to a number of patents.

\acknowledgments

The authors thank Simone Giani for valuable discussions.
Berkan Kaynak and Onur Potok contributed to the data collection from the Web 
of Science.


\bibliographystyle{JHEP}
\Urlmuskip=0mu plus 1mu\relax
\bibliography{biblio}

\providecommand{\href}[2]{#2}\begingroup\raggedright\begin{thebibliography}{10}

\bibitem{g4nim}
S.~Agostinelli et~al., \emph{Geant4 a simulation toolkit}, {\emph{Nucl.
  Instrum. Meth. A} {\bfseries 506} (2003) 250--303}.

\bibitem{g4tns}
J.~Allison et~al., \emph{Geant4 developments and applications}, {\emph{IEEE
  Trans. Nucl. Sci.} {\bfseries 53} (2006) 270--278}.

\bibitem{g4nim2}
J.~Allison et~al., \emph{Recent developments in {Geant4}}, {\emph{Nucl.
  Instrum. Meth. A} {\bfseries 835} (2016) 186--225}.

\bibitem{wos}
Clarivate, \emph{{The Web of Science\textsuperscript{\texttrademark}}},
  September, 2023.

\bibitem{P58}
A.~Dellacqua et~al., \emph{{GEANT4: an Object-Oriented toolkit for simulation
  in HEP}},  Tech. Rep. CERN/DRDC/P58, CERN, Geneva, Switzerland, 1994.

\bibitem{takaiwa_1993}
Y.~Takaiwa et~al., \emph{{Towards Object-Oriented GEANT - ProdiG Project}},  in
  \emph{{MC 93 - Proc. Int. Conf. Monte Carlo Simulation in High Energy and
  Nuclear Physics}}, pp.~339--350, World Scientific, 1994.

\bibitem{giani_1993}
S.~Giani, ``{Investigation of a class hierarchy for GEANT}.'' {Presented at the
  Mini-workshop on Object-Oriented GEANT, CERN/CN/AS}, 24-27 Aug., 1993.

\bibitem{intent}
A.~Fesefeldt et~al., \emph{{Letter of Intent to the DRDC}},  Tech. Rep.
  CERN/DRDC/94-28, CERN, Geneva, Switzerland, 1994.

\bibitem{rd44-97}
S.~Giani, \emph{{GEANT4: An Obiect-Oriented Toolkit for Simulation in HEP}},
  Tech. Rep. CERN/LHCC97-40, CERN, Geneva, Switzerland, 1997.

\bibitem{rd44-98}
S.~Giani, \emph{{GEANT4: An Object-Oriented Toolkit for Simulation in HEP}},
  Tech. Rep. CERN/LHCC98-44, CERN, Geneva, Switzerland, 1998.

\bibitem{higgs_atlas}
G.~Aad et~al., \emph{{Observation of a new particle in the search for the
  Standard Model Higgs boson with the ATLAS detector at the LHC}}, {\emph{Phys.
  Lett. B} {\bfseries 716} (2012) 1--29}.

\bibitem{higgs_cms}
S.~Chatrchyan et~al., \emph{{Observation of a new boson at a mass of 125 GeV
  with the CMS experiment at the LHC}}, {\emph{Phys. Lett. B} {\bfseries 716}
  (2012) 30--61}.

\bibitem{mann_1945}
H.~B. Mann, \emph{Nonparametric test against trend}, {\emph{Econometrica}
  {\bfseries 13} (1945) 245--259}.

\bibitem{kendall_1948}
M.~G. Kendall, \emph{Rank Correlation Methods}.
\newblock Griffin, 1948.

\bibitem{cox_1955}
D.~R. Cox and A.~Stuart, \emph{Some quick sign tests for trend in location and
  dispersion}, {\emph{Biometrika} {\bfseries 42} (1955) 80--95}.

\bibitem{R}
R.~Foundation, ``The {R} {P}roject for {S}tatistical {C}omputing.''
  https://www.r-project.org/.

\bibitem{egsnrcJournal}
I.~Kawrakow, \emph{{Accurate condensed history Monte Carlo simulation of
  electron transport. I. EGSnrc, the new EGS4 version}}, {\emph{Med. Phys.}
  {\bfseries 27} (2000) 485--498}.

\bibitem{fluka11}
T.~B{\"o}hlen et~al., \emph{{The FLUKA code: developments and challenges for
  high energy and medical applications}}, {\emph{Nucl. Data Sheets} {\bfseries
  120} (2014) 211--214}.

\bibitem{its3}
J.~A. Halbleib et~al., \emph{{ITS: the integrated TIGER series of
  electron/photon transport codes-Version 3.0}}, {\emph{IEEE Trans. Nucl. Sci.}
  {\bfseries 39} (1992) 1025--1030}.

\bibitem{mars}
N.~V. Mokhov et~al., \emph{{Recent enhancements to the MARS15 code}},
  {\emph{Radiat. Prot. Dosim.} {\bfseries 116} (2005) 99--103}.

\bibitem{mcnp6}
T.~Goorley et~al., \emph{Initial {MCNP6} release overview}, {\emph{Nucl.
  Technol.} {\bfseries 180} (2012) 298--315}.

\bibitem{openmc1}
P.~K. Romano and B.~Forget, \emph{{The OpenMC Monte Carlo particle transport
  code}}, {\emph{Ann. Nucl. Energy} {\bfseries 51} (2013) 274--281}.

\bibitem{penelope1995}
J.~Baro et~al., \emph{{PENELOPE: An algorithm for Monte Carlo simulation of the
  penetration and energy loss of electrons and positrons in matter}},
  {\emph{Nucl. Instrum. Meth. B} {\bfseries 100} (1995) 31--46}.

\bibitem{phits2002}
H.~Iwase et~al., \emph{Development of general-purpose particle and heavy ion
  transport {Monte Carlo} code}, {\emph{J. Nucl. Sci. Technol.} {\bfseries 39}
  (2002) 1142--1151}.

\bibitem{serpent2013}
J.~Leppanen et~al., \emph{{The Serpent Monte Carlo code: Status, development
  and applications in 2013}}, {\emph{Ann. Nucl. Energy} {\bfseries 82} (2015)
  142--150}.

\bibitem{tripoli4}
E.~Brun et~al., \emph{{TRIPOLI-4, CEA, EDF and AREVA reference Monte Carlo
  code}}, {\emph{Ann. Nucl. Energy} {\bfseries 82} (2015) 151--160}.

\bibitem{basaglia_2017}
T.~Basaglia, Z.~W. Bell, A.~Burger, P.~V. Dressendorfer and M.~G. Pia,
  \emph{Ghost science},  in \emph{2017 IEEE Nucl. Sci. Symp. Medical Imaging
  Conf.}, pp.~1--2, 2017.

\bibitem{shelx}
G.~M. Sheldrick, \emph{{A short history of SHELX}}, {\emph{Acta Crystallogr. A}
  {\bfseries 64} (2008) 112--122}.

\bibitem{qespresso}
P.~P.~Giannozzi et~al., \emph{{QUANTUM ESPRESSO: a modular and open-source
  software project for quantum simulations of materials}}, {\emph{J. Phys.
  Condens. Mat.} {\bfseries 21} (2009) 395502}.

\bibitem{pia_2014}
M.~G. Pia, T.~Basaglia, Z.~W. Bell and P.~V. Dressendorfer, \emph{Scholarly
  literature and the press: scientific impact and social perception of physics
  computing}, {\emph{J. Phys. Conf. Ser.} {\bfseries 513} (2014) 062039}.

\bibitem{vannoorden_2014}
R.~Van~Noorden, B.~Maher and R.~Nuzzo, \emph{The top 100 papers}, {\emph{Nature
  News} {\bfseries 514} (2014) 550}.

\bibitem{ineq_handbook}
F.~A. Cowell, \emph{Measurement of inequality},  in \emph{Handbook of Income
  Distribution}.
\newblock North Holland, 2015.

\bibitem{gini_1912}
C.~Gini, \emph{Variabilit{\`a} e mutabilit{\`a}: contributo allo studio delle
  distribuzioni e delle relazioni statistiche. [Fasc. I.]}.
\newblock Univ. of Cagliari, Tipogr. P. Cuppini, 1912.

\bibitem{pietra_1915}
G.~Pietra, \emph{Delle relazioni tra gli indici di variabilit{\`a} (nota i)},
  {\emph{Atti del Reale Istituto Veneto di Scienze, Lettere ed Arti, tomo
  LXXIV, parte II} (1915) 775--792}.

\bibitem{atkinson_1970}
A.~B. Atkinson, \emph{On the measurement of inequality}, {\emph{J. Econom.
  Theory} {\bfseries 2} (1970) 244 -- 263}.

\bibitem{hill_1973}
M.~O. Hill, \emph{Diversity and evenness: A unifying notation and its
  consequences}, {\emph{Ecology} {\bfseries 54} (1973) 427--432}.

\bibitem{roswell_2021}
M.~Roswell et~al., \emph{A conceptual guide to measuring species diversity},
  {\emph{Oikos} {\bfseries 130} (2021) 321--338}.

\bibitem{magurran_2010}
A.~E. Magurran and B.~J. McGill, \emph{Biological diversity: frontiers in
  measurement and assessment}.
\newblock OUP Oxford, 2010.

\bibitem{shannon_1948}
C.~E. Shannon, \emph{A mathematical theory of communication}, {\emph{Bell
  System Tech. J.} {\bfseries 27} (1948) 379--423}.

\bibitem{simpson_1949}
E.~H. Simpson, \emph{Measurement of diversity}, {\emph{Nature} {\bfseries 163}
  (1949) 688--688}.

\end{thebibliography}\endgroup



\end{document}